\documentclass[aps,reprint,onecolumn,notitlepage,11pt]{revtex4-1}
\usepackage{amsmath,amssymb,graphicx,subfigure,color}
\usepackage[hidelinks]{hyperref}

\begin{document}


\title{On the distribution and swim pressure of run-and-tumble \\ particles in confinement}

\author{Barath Ezhilan}
\author{Roberto Alonso-Matilla}
\author{David Saintillan}
\affiliation{Department of Mechanical and Aerospace Engineering, University of California San Diego, La Jolla, CA 92093, USA}

\date{\today}

\begin{abstract}
The spatial and orientational distribution in a dilute active suspension of non-Brownian run-and-tumble spherical swimmers confined between two planar hard walls is calculated theoretically. Using a kinetic model based on coupled bulk/surface probability density functions, we demonstrate the existence of a concentration wall boundary layer with thickness scaling with the run length, the absence of polarization throughout the channel, and the presence of sharp discontinuities in the bulk orientation distribution in the neighborhood of orientations parallel to the wall in the near-wall region. Our model is also applied to calculate the swim pressure in the system, which approaches the previously proposed ideal-gas behavior in wide channels but is found to decrease in narrow channels as a result of confinement. Monte-Carlo simulations are also performed for validation and show excellent quantitative agreement with our theoretical predictions.
 \end{abstract}

\maketitle

\section{Introduction\label{sec:introduction}}

The propensity of confined self-propelled particles to accumulate at boundaries is a trademark of active matter and has been reported in many experiments on bacterial suspensions \citep{Berke08,Gachelin13,Figueroa15} as well as simulations based on various models \citep{Hernandez05,Elgeti13,Li14}. Several disparate mechanisms have been proposed in explanation, including wall hydrodynamic interactions \citep{Berke08} and scattering due to collisions with the walls \citep{Li11}, though recent theoretical efforts have shown that the mere interplay of self-propulsion, stochastic processes and confinement is sufficient to explain accumulation \citep{Lee2013,Elgeti15,Ezhilan15}. With few exceptions, however, these models have necessitated particle diffusion, which in reality is nearly negligible in bacterial suspensions where stochasticity in the dynamics takes instead the form of run-and-tumble random walks \citep{Berg93}.

Understanding the distribution of active particles in confinement is especially critical for determining the mechanical force per unit area exerted by the suspension on the boundaries, or so-called `swim pressure'. This novel concept, which has received much scrutiny recently, describes the entropic force that must be applied on containing osmotic walls to keep self-propelled particles confined. Models based on the virial theorem \citep{Takatori14,Yang14,Winkler15} and on direct calculations of the wall mechanical pressure \citep{Solon15} in infinite or semi-infinite collections of spherical swimmers have all arrived at a simple ideal-gas law $\mathrm{\Pi}_{i}$ for the swim pressure in the limit of infinite dilution:  
\begin{equation}
\mathrm{\Pi}_{i}=n \zeta D_{t}=n\zeta \frac{V_{0}^{2}}{3\lambda}, \label{eq:idealpressure}
\end{equation}
where $n$ is the mean number density, $\zeta$ is the viscous drag coefficient of a particle and $D_{t}=V_{0}^{2}/3\lambda$ is the long-time translational diffusivity of an unconfined run-and-tumble swimmer expressed in terms of its speed $V_{0}$ and mean tumbling rate $\lambda$ \citep{Berg93}. Equation (\ref{eq:idealpressure}) and its extension to finite concentrations have proven useful to explain motility-induced phase separation in suspensions of self-propelled colloids \citep{Takatori14,Takatori15}, though its general validity as a thermodynamic equation of state for the pressure of active matter remains controversial \citep{Ray14,Mallory14,Ginot15} and appears to be limited to unconfined spherical particles \citep{Yang14,Solon15,Solon15b}. 

In this work, we analyze the simple case of a dilute suspension of athermal run-and-tumble spherical swimmers confined between two parallel flat plates. We propose in \S II a kinetic model based on two probability density functions describing the spatial and orientational distribution of the particles inside the gap and at the walls, which are coupled via flux conditions and only account for the effects of swimming and orientation decorrelation by tumbling. Further, our model implicitly captures hard-wall steric interactions without requiring the use of a soft potential to describe wall collisions as in previous theories \citep{Solon15,Solon15b}. A semi-analytical solution method is outlined in \S III, which provides the full probability density functions and allows for a direct calculation of the mechanical swim pressure exerted on the walls in terms of the polarization of the surface distributions. Results for the distributions and swim pressure are presented in \S IV, where they are shown to compare very favorably with Monte-Carlo simulations.

\section{Problem definition and theoretical model\label{sec:formulation}}

\subsection{Problem formulation}

\begin{figure}
\begin{center}
\hspace{-1cm}\includegraphics[width=10cm]{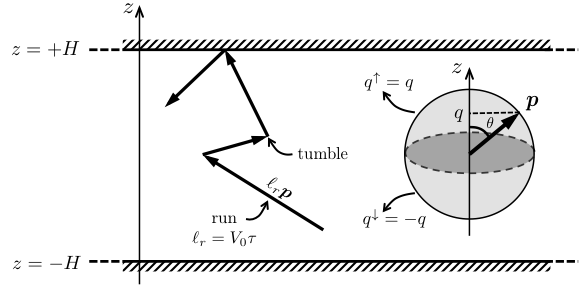}
\end{center}
\caption{Problem definition: run-and-tumble particles are confined between two flat plates separated by $2H$. The distribution of particles is a function of $z$ and $q=\boldsymbol{p}\boldsymbol{\cdot}\hat{\boldsymbol{z}}=\cos\theta\in(-1,1)$.  Orientations pointing towards the top and bottom walls are parametrized by $q^{\uparrow}=q$ and $q^{\downarrow}=-q$, respectively, both defined in $(0,1)$. } \label{fig:geometry}
\end{figure}

As a minimal model for an active suspension in confinement, we consider a dilute collection of self-propelled spherical particles confined between two infinite parallel plates separated by a distance $2H$ (see figure \ref{fig:geometry}). The swimmers are non-Brownian and simply perform a run-and-tumble random walk: straight runs of duration $\tau$ at constant velocity $V_{0}$ along the unit director $\boldsymbol{p}$ alternate with instantaneous tumbling events causing random and uncorrelated reorientations of  $\boldsymbol{p}$.   The time $\tau$ between tumbles is an exponentially distributed random variate with mean $\lambda^{-1}$, where the tumbling rate $\lambda$ is assumed to be independent of position and orientation.  To elucidate the interplay between run-and-tumble dynamics and confinement, we focus on the dilute limit and entirely neglect interparticle interactions. Particle-wall interactions are purely steric: as a swimmer meets one of the two surfaces, the normal component of its swimming motion is cancelled by a hard-core repulsive force causing it to stay at and push against the wall until a subsequent tumbling event reorients it into the bulk. Tumbling events occurring at the walls can lead to reorientation into the wall or into the bulk, so that a particle at a surface may need to undergo several tumbles before it is able to escape. 

There are only two  length scales in the problem: the mean run length $\ell_{r}=V_{0}\lambda^{-1}$ and the channel width $2H$. We define their ratio as the P\'eclet number $Pe=\ell_{r}/2H=V_{0}/2\lambda H$, where the two limits $Pe\rightarrow 0$ and $Pe\rightarrow \infty$ describe weak and strong confinement, respectively. Due to the symmetry of the problem, the distribution of particles in the channel only depends on two degrees of freedom: the wall-normal coordinate $z \in \left(-H,H \right)$ and the wall-normal component of the particle director $q=\boldsymbol{p}\boldsymbol{\cdot}\hat{\boldsymbol{z}} =\cos\theta \in \left( -1,1 \right)$. It is convenient to distinguish particles pointing towards the top and bottom walls, and to this end we divide the unit sphere of orientations into two hemispheres and define two distinct orientation coordinates $q^{\uparrow}=q\in(0,1)$ and $q^{\downarrow}=-q\in(0,1)$ on each hemisphere for particles pointing up or down, respectively, as depicted in figure~\ref{fig:geometry}. 

The distribution of particles in the channel is then fully described by a bulk probability density function $\psi(z,q)$ and by two surface probability density functions $\psi^{\uparrow}_{s}(q^{\uparrow})$ and $\psi_{s}^{\downarrow}(q^{\downarrow})$, which are only defined over half of the orientations since the surfaces cannot sustain a concentration of particles pointing towards the bulk. By symmetry, we expect
\begin{equation}
\psi(z,-q)=\psi(-z,q), \quad \psi(z,q^{\uparrow})=\psi(-z,q^{\downarrow}) \quad \mbox{and} \quad \psi_{s}^{\uparrow}(q^{\uparrow})=\psi_{s}^{\downarrow}(q^{\downarrow})
\end{equation}
for $q^{\uparrow}=q^{\downarrow}$.
Next, we describe the coupled bulk/surface conservation equations satisfied by these distributions, together with the appropriate boundary conditions.
 
\subsection{Bulk conservation equation} 

The steady bulk probability density function $\psi(z,q)$ satisfies the conservation equation
\begin{equation}
V_{0}\, q\, \frac{\partial}{\partial z} \psi\left(z,q\right) = -\lambda\,\psi\left(z,q\right) + \frac{1}{2} \int_{-1}^{1}\lambda\, \psi\left(z,q'\right)\,\mathrm{d}q'. \label{eq:bulkplane}
\end{equation}
The left-hand side describes transport along $z$ due to self-propulsion. Run-and-tumble dynamics is captured by the right-hand side, where the first term accounts for depletion due to swimmers tumbling away from orientation $q$, and the second term  for restoration due to swimmers tumbling from orientations $q'$ into  $q$.  
It is also useful to define the orientational moments of order $j$ of the bulk probability density function on the full sphere and on the upper/lower hemispheres of orientations as
\begin{equation}
 M_{j}(z) = \int_{-1}^{1} q^{j}\,\psi\left(z,q\right) \, \mathrm{d}q \quad \mbox{and} \quad
 M_{j}^{\uparrow\downarrow}(z) = \int_{0}^{1} (q^{\uparrow\downarrow})^{j}\,\psi (z,q^{\uparrow\downarrow}) \, \mathrm{d}q^{\uparrow\downarrow},
\end{equation}
and we note that the zeroth, first and second moments correspond to the concentration, polarization, and nematic order parameter fields:
\begin{eqnarray}
 c(z) = M_{0}(z), \quad \quad &m(z) = M_{1}(z), \quad \quad &S(z) = M_{2}(z), \\
 c^{\uparrow\downarrow}(z) = M_{0}^{\uparrow\downarrow}(z), \quad \quad &m^{\uparrow\downarrow}(z) = M_{1}^{\uparrow\downarrow}(z), \quad \quad &S^{\uparrow\downarrow}(z) = M_{2}^{\uparrow\downarrow}(z).
\end{eqnarray}
By symmetry, it is straightforward to see that full moments of even order are even functions of $z$ whereas those of odd order are odd functions. With these notations, the bulk conservation equation (\ref{eq:bulkplane}) simplifies to 
\begin{equation}
\ell_{r}\, q\, \frac{\partial}{\partial z} \psi\left(z,q\right) = - \psi(z,q)+\tfrac{1}{2}c(z).  \label{eq:bulkPDF}
\end{equation}

\subsection{Surface conservation equations}

Similarly, conservation equations for the steady surface probability density functions at the walls can be written. We first define the surface concentration and polarization as
\begin{equation}
c_{s}=\int_{0}^{1}\psi_{s}^{\uparrow\downarrow}(q^{\uparrow\downarrow})\,\mathrm{d}q^{\uparrow\downarrow} \quad \mbox{and} \quad m_{s}=\int_{0}^{1} q^{\uparrow\downarrow}\,  \psi_{s}^{\uparrow\downarrow}(q^{\uparrow\downarrow})\,\mathrm{d}q^{\uparrow\downarrow} ,
\end{equation}
and note that the values of $c_{s}$ and $m_{s}$ are the same at both walls. With these notations, the conservation equation at the upper wall  ($z=+H$) reads
\begin{equation}
V_{0} \,q^{\uparrow}\, \psi(H ,q^{\uparrow}) = \lambda \left[ \psi_{s}^{\uparrow} (q^{\uparrow}) - \tfrac{1}{2} c_{s} \right],\label{eq:surfacePDF}
\end{equation}
and a similar equation holds at $z=-H$. The right-hand side in equation (\ref{eq:surfacePDF}) describes tumbling processes at the wall. The left-hand side, on the other hand, captures the flux of particles that enter the surface from the bulk by self-propulsion, and is therefore proportional to the bulk probability density function $\psi(H ,q^{\uparrow})$ next to the wall. Evaluating the zeroth and first orientational moments of equation (\ref{eq:surfacePDF}) yields simple relations between $c_{s}$ and $m_{s}$ and the values of the bulk moments in the vicinity of the wall:
\begin{equation}
c_{s}=2\ell_{r}m^{\uparrow}(H), \quad \quad m_{s}=\ell_{r}\left[\tfrac{1}{2}m^{\uparrow}(H)+S^{\uparrow}(H)\right].  \label{eq:surfacerelations}
\end{equation}

\subsection{Boundary condition and particle number conservation}

Equation (\ref{eq:surfacePDF}) can be interpreted as a boundary condition for orientations pointing into the wall. For orientations pointing away from the wall, the swimming flux away from the wall must be balanced by tumbling of particles from the surface towards the bulk. Simply stated, particles on the surface that tumble to an orientation pointing into the bulk are transported away by self-propulsion. This leads to the additional condition 
\begin{equation}
V_{0}\,q^{\downarrow}\,\psi(H,q^{\downarrow})=\tfrac{1}{2}\lambda\, c_{s} \quad \quad \mbox{or} \quad \quad  \ell_{r}\,q^{\downarrow}\,\psi(H,q^{\downarrow})=\tfrac{1}{2}c_{s}. \label{eq:BC}
\end{equation}
As $c_{s}$ is constant and finite, this condition suggests divergence and discontinuity of the bulk probability density function for orientations parallel to the wall  ($q^{\downarrow}\rightarrow 0$), as will indeed be verified in our analytical solution and stochastic simulations. 

Finally, the above system of equations for the bulk and surface distributions is supplemented by a constraint on the total number of particles in the channel:
\begin{equation}
2\,c_{s}+\int_{-H}^{H}c(z)\,\mathrm{d}z=N,
\end{equation}
where $N$ is the total particle number in a vertical slice of unit horizontal cross-section.

\section{Method of solution and swim pressure calculation\label{sec:solutionmethod}}

\subsection{Integral equation for the moments}

We now outline a solution method for the system described in \S \ref{sec:formulation}. As a first step, we derive an integral equation relating the bulk orientational moments to the concentration field. The bulk concentration equation (\ref{eq:bulkPDF}) can be viewed as a linear inhomogeneous ordinary differential equation for $\psi(z,q)$ where $q$ is a parameter. We solve it by the method of variation of constants, treating orientations $q^{\uparrow}$ and $q^{\downarrow}$ separately. After applying the boundary conditions (\ref{eq:surfacePDF}) and (\ref{eq:BC}), we obtain a general expression for the bulk probability density function:
\begin{equation}
 \psi(z,q^{\uparrow\downarrow})  = \frac{c_{s}}{2\ell_{r}\,q^{\uparrow\downarrow}}\exp\left[-\frac{\left(H \pm z\right)}{\ell_{r}\,q^{\uparrow\downarrow}}\right] \pm \int_{\mp H}^{z} \frac{c(z')}{2 \ell_{r}\,q^{\uparrow\downarrow}} \exp\left[\mp\frac{(z-z')}{\ell_{r}\,q^{\uparrow\downarrow}}\right]\,\mathrm{d}z'.\label{eq:psi_sol}
\end{equation}
Note that the bulk and surface concentrations $c(z)$ and $c_{s}$ both appear on the right-hand side and are still unknown. However, equation (\ref{eq:psi_sol}) shows that their knowledge entirely specifies the bulk distribution $\psi(z,q)$. The bulk moments of order $j$ on both hemispheres of orientations are immediately obtained by integration:
\begin{equation}
M_{j}^{\uparrow \downarrow}(z) = \frac{c_{s}}{2\ell_{r}}\mathcal{E}_{j+1}\left[\frac{H \pm z}{\ell_{r}}\right] \pm \int_{\mp H}^{z} \frac{c(z')}{2\ell_{r}}\,\mathcal{E}_{j+1}\left[\pm\frac{(z-z')}{\ell_{r}}\right]\,\mathrm{d}z',
\end{equation}
where $\mathcal{E}_{j}$ is the exponential integral function defined as
\begin{equation}
\mathcal{E}_{j}(z) = \int_{0}^{1} u^{j-2}\exp\left(-\frac{z}{u}\right) \,\mathrm{d}u.
\end{equation}
Finally, the moment of order $j$ on the full sphere of orientations can be shown to be
\begin{equation}
M_{j}(z) = \frac{c_{s}}{2\ell_{r}}\left(\mathcal{E}_{j+1}\left[\frac{H + z}{\ell_{r}}\right] + \mathcal{E}_{j+1}\left[\frac{H - z}{\ell_{r}}\right]\right) + \int_{-H}^{H} \frac{c(z')}{2\ell_{r}}\mathcal{E}_{j+1}\left[\left|\frac{z-z'}{\ell_{r}}\right|\right]\,\mathrm{d}z'.\label{eq:moment_sol}
\end{equation}

\subsection{Bulk concentration profile}

Setting $j=0$ in equation (\ref{eq:moment_sol}) immediately provides an integral equation for the yet unknown concentration profile:
\begin{equation}
c(z) = \frac{c_{s}}{2\ell_{r}}\left(\mathcal{E}_{1}\left[\frac{H + z}{\ell_{r}}\right] + \mathcal{E}_{1}\left[\frac{H - z}{\ell_{r}}\right]\right) + \int_{-H}^{H} \frac{c(z')}{2\ell_{r}}\mathcal{E}_{1}\left[\left|\frac{z-z'}{\ell_{r}}\right|\right]\,\mathrm{d}z'. \label{eq:concentration_sol}
\end{equation}
Dividing through by $c_{s}$, we obtain an equation for $c(z)/c_{s}$ that can be solved numerically. For finite $\ell_{r}$, we find that an approximate solution is easily obtained iteratively by casting equation (\ref{eq:concentration_sol}) in the form $c_{k+1}(z)/c_{s}=f[c_{k}(z)/c_{s}]$, starting with an initial guess which we take to be $c_{0}(z)=0$. In strong confinement (large $Pe$), the solution converges in $O(20)$ iterations, though more iterations are required in wider channels.

\subsection{Surface concentration}

To complete the solution, the value of the surface concentration $c_{s}$ must be calculated. To this end, we make use of a crucial property of the system, namely the overall isotropy of the suspension. Indeed, the spatially averaged orientation distribution $\mathcal{Q}(q)$ must be isotropic as reorientation due to tumbling is completely uncorrelated and is unaffected by the presence of the walls. This is expressed mathematically as
\begin{equation}
\mathcal{Q}(q)=\psi_{s}^{\uparrow\downarrow}(q^{\uparrow\downarrow}) +\int_{-H}^{H}\psi(z,q)\,\mathrm{d}z=\frac{N}{2}, \label{eq:overallisotropy1}
\end{equation}
which can be combined with the surface conservation equation (\ref{eq:surfacePDF}) to provide an equation for $c_{s}$. The solution to the problem then proceeds as follows. Solving equation (\ref{eq:concentration_sol}) using the iterative procedure outlined above provides a solution for $c(z)/c_{s}$. This can be inserted in equation (\ref{eq:psi_sol}) to obtain $\psi(z,q)/c_{s}$, which can then be substituted into the overall isotropy condition (\ref{eq:overallisotropy1}) to solve for $c_{s}$. As a final step, the surface probability density function $\psi_{s}$ can be determined using equation (\ref{eq:surfacePDF}). Solutions obtained by this method are presented in \S \ref{sec:results}, where excellent agreement with results from Monte-Carlo simulations will be shown.

\subsection{Swim pressure calculation}

The above formulation  provides a direct way of estimating the swim pressure in the system, which is simply the force per unit area exerted by the particles at the walls as they push on the surface. Specifically, the normal component of the motion of each particle at the upper wall is resisted by a force  $\zeta V_{0} q^{\uparrow}$, where $\zeta$ is the viscous drag coefficient of one particle \citep{Takatori14}. Knowing the surface probability density function $\psi_{s}^{\uparrow}$, an expression for the swim pressure is then easily found as
\begin{equation}
\mathrm{\Pi}_{s}=\int_{0}^{1}\zeta V_{0}q^{\uparrow}\psi_{s}^{\uparrow}(q^{\uparrow})\,\mathrm{d}q^{\uparrow}=\zeta V_{0} m_{s},
\end{equation}
where $m_{s}$ is the surface polarization. Using equation (\ref{eq:surfacerelations}), this is also expressed in terms of bulk variables as
\begin{equation}
\mathrm{\Pi}_{s}=\zeta V_{0} \ell_{r}\left[\tfrac{1}{2}m^{\uparrow\downarrow}(\pm H)+S^{\uparrow\downarrow}(\pm H)\right]=\zeta \frac{V_{0}^{2}}{\lambda}\left[\tfrac{1}{2}m^{\uparrow\downarrow}(\pm H)+S^{\uparrow\downarrow}(\pm H)\right]. \label{eq:pressure}
\end{equation}
In bulk unconfined systems, previous models have led to the ideal-gas pressure $\mathrm{\Pi}_{i}$ of equation~(\ref{eq:idealpressure}),
which contains no information on particle orientations due to isotropy but follows the same scaling as equation (\ref{eq:pressure}). To compare both predictions, we define a dimensionless pressure as the ratio of equations (\ref{eq:pressure}) and (\ref{eq:idealpressure}):
\begin{equation}
\mathcal{P}=\frac{\mathrm{\Pi}_{s}}{\mathrm{\Pi}_{i}}=\frac{3m_{s}}{n\ell_{r}}=\frac{3}{n}\left[\tfrac{1}{2}m^{\uparrow\downarrow}(\pm H)+S^{\uparrow\downarrow}(\pm H)\right],
\end{equation}
where $n=N/2H$ is the mean number density in our system. $\mathcal{P}-1$ quantifies the departure from the ideal-gas swim pressure. We will see in \S 4 that $\mathcal{P}\rightarrow 1$ in very wide channels ($Pe\rightarrow 0$), but deviates from $1$ when $Pe>0$ as a result of confinement.

\section{Results and comparison to simulations\label{sec:results}}

\subsection{Simulation method}

To validate our  model, we also perform Markov-chain Monte-Carlo simulations of run-and-tumble swimmers between two hard walls. During a run of duration $\tau$, the swimmer trajectory simply evolves as $\boldsymbol{x}(t+\mathrm{\Delta} t)=\boldsymbol{x}(t)+ V_{0} \boldsymbol{p}\,\mathrm{\Delta} t$ where $\mathrm{\Delta} t$ is a short time step. Each run is then followed by a tumbling event, where the new orientation vector $\boldsymbol{p}$ is picked randomly on the unit sphere. The time $\tau$ between two consecutive tumbles is drawn from an exponential distribution with cumulative distribution function $F(\tau)=1-\exp[{-\lambda \tau}]$. When a swimmer meets a wall, it remains there and continues to tumble until it reorients towards the bulk and swims away. Time-averaged bulk and surface probability density functions were extracted from orientational and spatial histograms, and convergence was checked with respect to $\mathrm{\Delta}t$ and to the duration of the simulation.

\subsection{Theoretical and numerical results}

\begin{figure}
\centering
  \includegraphics[scale=0.95]{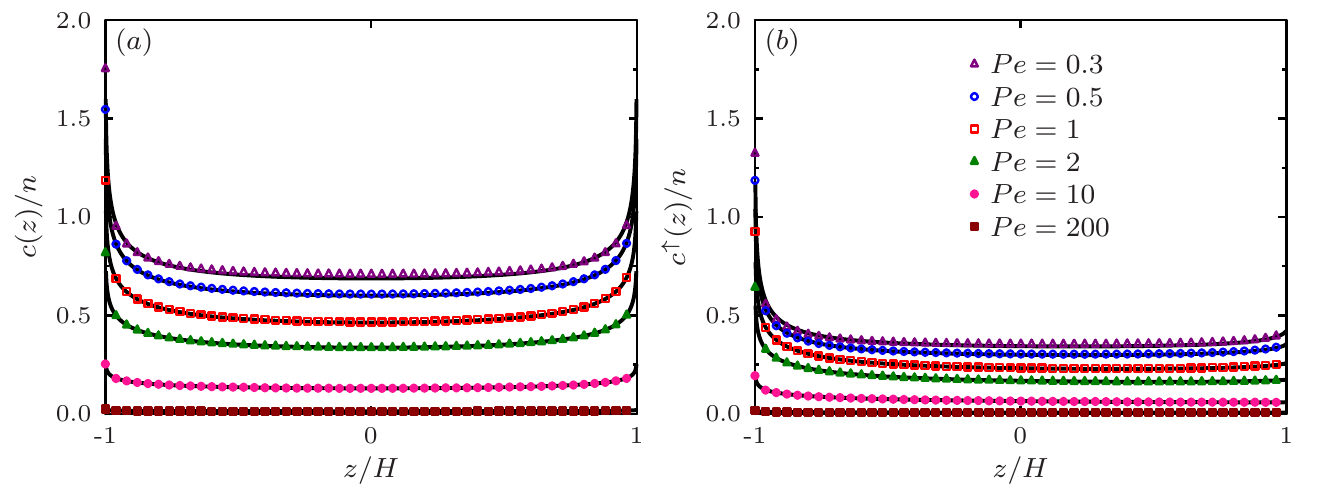}
\caption{Concentration profiles across the channel for various values of $Pe=\ell_{r}/2H$: ($a$) full concentration $c(z)$, and ($b$) partial `up' concentration $c^{\uparrow}(z)$. Solid lines show the semi-analytical solution of \S \ref{sec:solutionmethod}, and symbols are Monte-Carlo simulation results.}
  \label{fig:bulkconcentration}
\end{figure}

Solutions for the bulk concentration profile are depicted in figure~\ref{fig:bulkconcentration}, where both the full concentration $c(z)$ and the partial `up' concentration $c^{\uparrow}(z)$ are plotted for various values of the P\'eclet number, which measures the degree of confinement. The full concentration profiles in figure~\ref{fig:bulkconcentration}($a$) show significant accumulation at the walls, with wall boundary layers whose thickness scales with $\ell_{r}$. An interesting and unique feature of run-and-tumble particles is that accumulation occurs in the absence of  polarization, and $m(z)$ is found to be strictly zero throughout the channel (not shown). A non-zero polarization would indeed lead to a net flux of particles in the wall-normal direction, which cannot happen in a confined athermal system, unlike in Brownian suspensions where this flux can be balanced by diffusion \citep{Ezhilan15}. In fact, averaging equation~(\ref{eq:bulkplane}) over $q$ immediately leads to the condition that $m(z)=0$. The profiles also show the presence of a singularity in $c(z)$ at the walls, which is a direct consequence of the boundary condition (\ref{eq:BC}) and is also obvious from the solution (\ref{eq:concentration_sol}) where $\mathcal{E}_{1}(0)$ diverges. Concentration singularities were also predicted by Elgeti \& Gompper \cite{Elgeti15}, though their model did not capture orientation distributions. As confinement becomes significant and $Pe$ increases, the bulk concentration decreases throughout the channel to reach nearly zero at $Pe=200$, indicating that strongly confined particles spend most of their time at the boundaries. Excellent quantitative agreement is obtained between theory and Monte-Carlo simulations, thereby strongly validating our kinetic  model. 

\begin{figure}
\centering
   \includegraphics[scale=0.95]{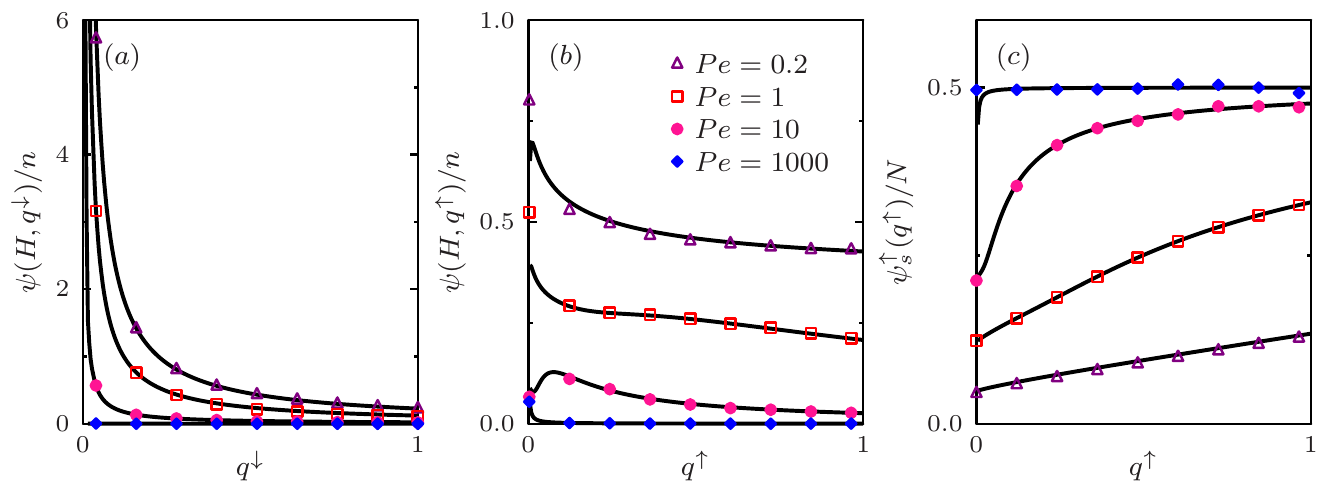}
\caption{Bulk probability density at the top wall for ($a$) orientations pointing away from the wall and  ($b$) orientations pointing towards it. ($c$) Surface probability density at the top wall as a function of $q^{\uparrow}$. Solid lines show the semi-analytical solution of \S \ref{sec:solutionmethod}, and symbols are Monte-Carlo simulation results.} 
  \label{fig:BCsurfacePDF}
\end{figure}

Figure~\ref{fig:bulkconcentration}($b$) also shows the partial `up' concentration obtained by only counting particles pointing towards the top wall. The asymmetry of the profiles and the singularity at the bottom wall indicates that on average there are more particles pointing away from the wall than towards it inside the wall accumulation layers. However, in order to satisfy no net polarization in the bulk, this implies that those particles pointing towards the wall are more strongly polarized than those pointing away. This point is confirmed in figure~\ref{fig:BCsurfacePDF}($a$--$b$), showing the orientation distributions in the bulk in the vicinity of the top wall for orientations pointing away from and towards the wall. Figure~\ref{fig:BCsurfacePDF}($a$) confirms the divergence of the bulk probability density in the neighborhood of orientations parallel to the wall ($q^{\downarrow}\rightarrow 0$) as expected from boundary condition (\ref{eq:BC}), which is also captured by the simulations. The presence of this discontinuity can be rationalized as follows: particles that leave the surface at an orientation $q^{\downarrow}\gtrsim 0$ swim nearly parallel to the surface and therefore remain there much longer than particles leaving in other orientations. The distribution of particles pointing towards the wall in figure~\ref{fig:BCsurfacePDF}($b$) shows no such singularity, but exhibits a finite peak at a critical value of $q^{\uparrow}$ whose origin remains unclear. The orientation distribution $\psi_{s}^{\uparrow}(q^{\uparrow})$ of particles on the top wall is shown in figure~\ref{fig:BCsurfacePDF}($c$) and shows a preferential alignment normal to the wall rather than parallel to it. However, this distribution becomes nearly isotropic under very strong confinement ($Pe=1000$), for reasons that we elucidate below.

\begin{figure}
\centering
  \includegraphics[scale=0.95]{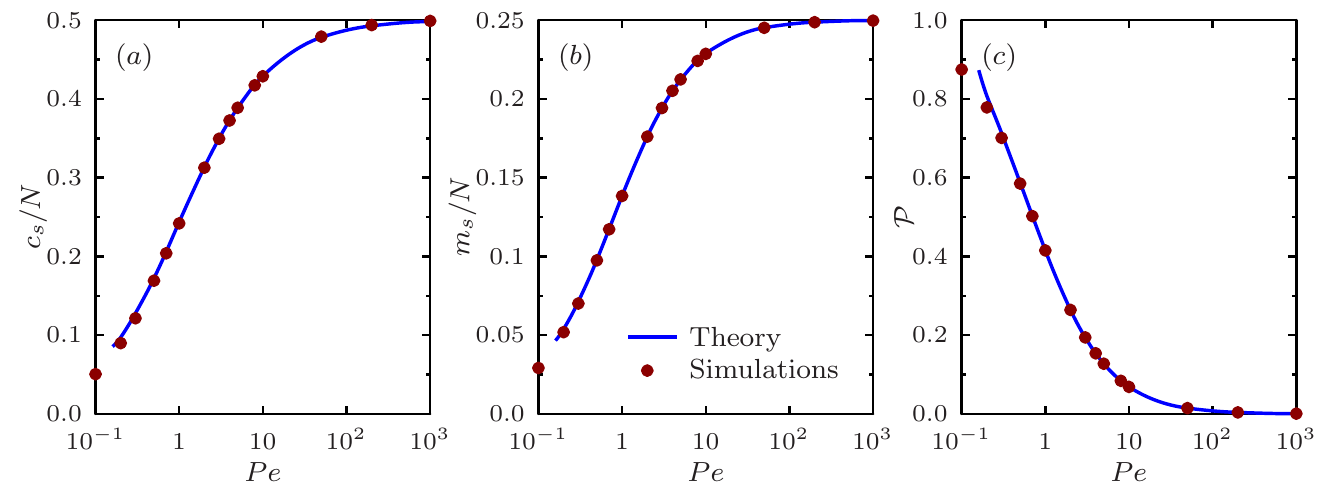}
\caption{($a$) Surface concentration $c_{s}$, ($b$) surface polarization $m_{s}$, and  ($c$) dimensionless pressure $\mathcal{P}$ as functions of P\'eclet number $Pe = \ell_{r}/2H$. Solid lines show the semi-analytical solution of \S \ref{sec:solutionmethod}, and symbols are Monte-Carlo simulation results.}
  \label{fig:surfacemoments}
\end{figure}

Taking moments of $\psi_{s}^{\uparrow}(q^{\uparrow})$ provides the surface concentration $c_{s}$ and surface polarization $m_{s}$, which are plotted versus P\'eclet number in figure~\ref{fig:surfacemoments}($a$--$b$). Both quantities increase with increasing confinement, but asymptote as $Pe\rightarrow \infty$. The asymptote for $c_{s}$ is $N/2$, meaning that in very narrow channels the particles spend all their time at the boundaries; indeed, the time $2H/V_{0}$ it takes them to cross the gap is infinitesimal compared to the mean run time $\lambda^{-1}$. This is also consistent with the decrease in the bulk concentration seen in figure~\ref{fig:bulkconcentration}($a$). In this limit, particles tumbling away from one wall reach the other wall nearly instantaneously, leading to an isotropic surface orientation distribution in agreement with figure~\ref{fig:BCsurfacePDF}($c$), hence the asymptote of $N/4$ for the wall polarization.

Lastly, the dependence of the dimensionless swim pressure $\mathcal{P}$ on the degree of confinement is illustrated in figure~\ref{fig:surfacemoments}($c$). In the limit of weak confinement ($H\gg \ell_{r}$ or $Pe\rightarrow 0$), the swim pressure is seen to tend to the ideal-gas law of equation (\ref{eq:idealpressure}) in both our model and simulations: $\mathcal{P}\rightarrow 1$ or $\mathrm{\Pi}_{s}\rightarrow \mathrm{\Pi}_{i}$. This corresponds to the limit of a single wall where the gap width $H$ plays no role, and validates the results of previous studies in infinite or semi-infinite systems for which the expression for $\mathrm{\Pi}_{i}$ was first derived \citep{Takatori14,Solon15}. Confinement, however, causes a decrease in the swim pressure, which in fact tends to zero for fixed $n$ in very narrow gaps. The high-$Pe$ asymptote for $m_{s}$ describes the limiting behavior:
\begin{equation}
\mathcal{P}\rightarrow \frac{3}{4}Pe^{-1}, \quad \mbox{i.e.} \quad \mathrm{\Pi}_{s}\rightarrow \frac{3}{4}Pe^{-1}\mathrm{\Pi}_{i}=\frac{nH\zeta V_{0}}{2}=\frac{N\zeta V_{0}}{4}
\end{equation}
as $Pe\rightarrow \infty$ (or $H\rightarrow 0$), which corresponds to $N/2$ particles pushing with an average force of $\zeta V_{0}/2$ against each wall. The decrease in pressure and the details of the asymptote agree with the previous two-dimensional results of   Yang \textit{et al.}~\cite{Yang14}, who also verified them in numerical simulations of self-propelled disks. They are also consistent with the study of Ray \textit{et al.}~\cite{Ray14}, who analyzed the force on two nearby parallel plates in an active particle bath and proposed that the pressure inside the gap in a one-dimensional system with constant run length goes as $\mathrm{\Pi}_{i}/(1+Pe)$. 

\subsection{Summary and discussion} 

We have presented a simple continuum model for a dilute suspension of spherical run-and-tumble particles confined between two hard walls and interacting via purely steric forces with the walls. The model improves upon our previous theory for confined Brownian suspensions \citep{Ezhilan15} by allowing us to address the limit of zero temperature for the first time within a continuum framework and by incorporating a more realistic treatment of surface interactions and exchange processes between surfaces and the bulk without the need for a soft potential \citep{Solon15}. This description also provides a direct and simple way of calculating the mechanical swim pressure exerted on the walls. We have outlined an elegant approach to derive a semi-analytical solution for the probability density functions, and demonstrated excellent quantitative agreement between our model and results from discrete Monte-Carlo simulations. 

Our theoretical predictions and simulation results have highlighted several striking features of confined suspensions of run-and-tumble particles, namely the presence of a singularity and discontinuity in the bulk probability density function for orientations nearly parallel to the walls in the near-wall region, and the existence of a concentration boundary layer of thickness of the order of $\ell_{r}$ that actually diverges at the walls. Our pressure calculations were shown to match the recently proposed ideal-gas equation of state of active matter in wide channels, thus further validating this ideal-gas law and confirming the prediction that the precise nature of particle-wall steric interactions has no impact on the wall mechanical pressure for spherical particles \citep{Solon15}. 
We demonstrated, however, that confinement leads to departures from this ideal behavior and specifically to a decrease in the swim pressure, which in fact vanishes in the limit of an infinitely narrow gap. In this case, we found that swimmers spend all their time at the boundaries, which provides the basis for previous models of strongly confined systems that only account for the surface distribution of swimmers \citep{Fily14}. 

While capturing the salient features of confined active suspensions, the problem under consideration remained minimal. Yet, the kinetic model presented here could be further modified to incorporate other effects and provide a more realistic description of biological or synthetic active systems. In particular, many active particles are rod-shaped and therefore also incur an aligning torque as they interact with boundaries. Recent theoretical work has shown that the wall pressure is modified in that case and becomes dependent upon the precise nature of particle-wall interactions \citep{Solon15b}. In addition, experiments show that the surface-to-bulk tumbling of biological swimmers as well as certain types of synthetic swimmers is not uncorrelated but rather results in the preferential release of the particles near a specific angle \citep{Kantsler13,Volpe11}. Incorporating such details in our model is straightforward and would modify the distribution of particles near the walls with unexpected consequences for the mechanical pressure. Our basic model, validated here in the dilute limit, could also be modified to account for hydrodynamic couplings and to study the structure  of the self-generated flows and collective dynamics of interacting active particles in confinement. Extending the model to non-planar boundaries, whether concave or convex, is not as straightforward but would be of great interest for the theoretical description of active particle transport in complex geometries or of their interaction with and transport of passive payloads. This rich avenue is the focus of our current work.

\begin{acknowledgments}
The authors thank John F. Brady for seminal discussions, and gratefully acknowledge funding from NSF Grants CBET-1532652 and DMS-1463965.
\end{acknowledgments}

\bibliography{RTbib}

\begin{thebibliography}{23}%
\makeatletter
\providecommand \@ifxundefined [1]{%
 \@ifx{#1\undefined}
}%
\providecommand \@ifnum [1]{%
 \ifnum #1\expandafter \@firstoftwo
 \else \expandafter \@secondoftwo
 \fi
}%
\providecommand \@ifx [1]{%
 \ifx #1\expandafter \@firstoftwo
 \else \expandafter \@secondoftwo
 \fi
}%
\providecommand \natexlab [1]{#1}%
\providecommand \enquote  [1]{``#1''}%
\providecommand \bibnamefont  [1]{#1}%
\providecommand \bibfnamefont [1]{#1}%
\providecommand \citenamefont [1]{#1}%
\providecommand \href@noop [0]{\@secondoftwo}%
\providecommand \href [0]{\begingroup \@sanitize@url \@href}%
\providecommand \@href[1]{\@@startlink{#1}\@@href}%
\providecommand \@@href[1]{\endgroup#1\@@endlink}%
\providecommand \@sanitize@url [0]{\catcode `\\12\catcode `\$12\catcode
  `\&12\catcode `\#12\catcode `\^12\catcode `\_12\catcode `\%12\relax}%
\providecommand \@@startlink[1]{}%
\providecommand \@@endlink[0]{}%
\providecommand \url  [0]{\begingroup\@sanitize@url \@url }%
\providecommand \@url [1]{\endgroup\@href {#1}{\urlprefix }}%
\providecommand \urlprefix  [0]{URL }%
\providecommand \Eprint [0]{\href }%
\providecommand \doibase [0]{http://dx.doi.org/}%
\providecommand \selectlanguage [0]{\@gobble}%
\providecommand \bibinfo  [0]{\@secondoftwo}%
\providecommand \bibfield  [0]{\@secondoftwo}%
\providecommand \translation [1]{[#1]}%
\providecommand \BibitemOpen [0]{}%
\providecommand \bibitemStop [0]{}%
\providecommand \bibitemNoStop [0]{.\EOS\space}%
\providecommand \EOS [0]{\spacefactor3000\relax}%
\providecommand \BibitemShut  [1]{\csname bibitem#1\endcsname}%
\let\auto@bib@innerbib\@empty
\bibitem [{\citenamefont {Berke}\ \emph {et~al.}(2008)\citenamefont {Berke},
  \citenamefont {Turner}, \citenamefont {Berg},\ and\ \citenamefont
  {Lauga}}]{Berke08}%
  \BibitemOpen
  \bibfield  {author} {\bibinfo {author} {\bibfnamefont {A.~P.}\ \bibnamefont
  {Berke}}, \bibinfo {author} {\bibfnamefont {L.}~\bibnamefont {Turner}},
  \bibinfo {author} {\bibfnamefont {H.~C.}\ \bibnamefont {Berg}}, \ and\
  \bibinfo {author} {\bibfnamefont {E.}~\bibnamefont {Lauga}},\ }\href@noop {}
  {\bibfield  {journal} {\bibinfo  {journal} {Phys. Rev. Lett.}\ }\textbf
  {\bibinfo {volume} {101}},\ \bibinfo {pages} {038102} (\bibinfo {year}
  {2008})}\BibitemShut {NoStop}%
\bibitem [{\citenamefont {Gachelin}\ \emph {et~al.}(2013)\citenamefont
  {Gachelin}, \citenamefont {{Mi\~no}}, \citenamefont {Berthet}, \citenamefont
  {Lindner}, \citenamefont {Rousselet},\ and\ \citenamefont
  {{Cl\'ement}}}]{Gachelin13}%
  \BibitemOpen
  \bibfield  {author} {\bibinfo {author} {\bibfnamefont {J.}~\bibnamefont
  {Gachelin}}, \bibinfo {author} {\bibfnamefont {G.}~\bibnamefont {{Mi\~no}}},
  \bibinfo {author} {\bibfnamefont {H.}~\bibnamefont {Berthet}}, \bibinfo
  {author} {\bibfnamefont {A.}~\bibnamefont {Lindner}}, \bibinfo {author}
  {\bibfnamefont {A.}~\bibnamefont {Rousselet}}, \ and\ \bibinfo {author}
  {\bibfnamefont {E.}~\bibnamefont {{Cl\'ement}}},\ }\href@noop {} {\bibfield
  {journal} {\bibinfo  {journal} {Phys. Rev. Lett.}\ }\textbf {\bibinfo
  {volume} {110}},\ \bibinfo {pages} {268103} (\bibinfo {year}
  {2013})}\BibitemShut {NoStop}%
\bibitem [{\citenamefont {{Figueroa-Morales}}\ \emph
  {et~al.}(2015)\citenamefont {{Figueroa-Morales}}, \citenamefont {{Mi\~no}},
  \citenamefont {Rivera}, \citenamefont {Caballero}, \citenamefont
  {{Cl\'ement}}, \citenamefont {Altshuler},\ and\ \citenamefont
  {Lindner}}]{Figueroa15}%
  \BibitemOpen
  \bibfield  {author} {\bibinfo {author} {\bibfnamefont {N.}~\bibnamefont
  {{Figueroa-Morales}}}, \bibinfo {author} {\bibfnamefont {G.}~\bibnamefont
  {{Mi\~no}}}, \bibinfo {author} {\bibfnamefont {A.}~\bibnamefont {Rivera}},
  \bibinfo {author} {\bibfnamefont {R.}~\bibnamefont {Caballero}}, \bibinfo
  {author} {\bibfnamefont {E.}~\bibnamefont {{Cl\'ement}}}, \bibinfo {author}
  {\bibfnamefont {E.}~\bibnamefont {Altshuler}}, \ and\ \bibinfo {author}
  {\bibfnamefont {A.}~\bibnamefont {Lindner}},\ }\href@noop {} {\bibfield
  {journal} {\bibinfo  {journal} {Soft Matter}\ ,\ \bibinfo {pages}
  {DOI:10.1039/C5SM00939A}} (\bibinfo {year} {2015})}\BibitemShut {NoStop}%
\bibitem [{\citenamefont {{Hern\'andez-Ortiz}}\ \emph
  {et~al.}(2005)\citenamefont {{Hern\'andez-Ortiz}}, \citenamefont {Stoltz},\
  and\ \citenamefont {Graham}}]{Hernandez05}%
  \BibitemOpen
  \bibfield  {author} {\bibinfo {author} {\bibfnamefont {J.~P.}\ \bibnamefont
  {{Hern\'andez-Ortiz}}}, \bibinfo {author} {\bibfnamefont {C.~G.}\
  \bibnamefont {Stoltz}}, \ and\ \bibinfo {author} {\bibfnamefont {M.~D.}\
  \bibnamefont {Graham}},\ }\href@noop {} {\bibfield  {journal} {\bibinfo
  {journal} {Phys. Rev. Lett.}\ }\textbf {\bibinfo {volume} {95}},\ \bibinfo
  {pages} {204501} (\bibinfo {year} {2005})}\BibitemShut {NoStop}%
\bibitem [{\citenamefont {Elgeti}\ and\ \citenamefont
  {Gompper}(2013)}]{Elgeti13}%
  \BibitemOpen
  \bibfield  {author} {\bibinfo {author} {\bibfnamefont {J.}~\bibnamefont
  {Elgeti}}\ and\ \bibinfo {author} {\bibfnamefont {G.}~\bibnamefont
  {Gompper}},\ }\href@noop {} {\bibfield  {journal} {\bibinfo  {journal}
  {Europhys. Lett.}\ }\textbf {\bibinfo {volume} {101}},\ \bibinfo {pages}
  {48003} (\bibinfo {year} {2013})}\BibitemShut {NoStop}%
\bibitem [{\citenamefont {Li}\ and\ \citenamefont {Ardekani}(2014)}]{Li14}%
  \BibitemOpen
  \bibfield  {author} {\bibinfo {author} {\bibfnamefont {G.}~\bibnamefont
  {Li}}\ and\ \bibinfo {author} {\bibfnamefont {A.~M.}\ \bibnamefont
  {Ardekani}},\ }\href@noop {} {\bibfield  {journal} {\bibinfo  {journal}
  {Phys. Rev. E}\ }\textbf {\bibinfo {volume} {90}},\ \bibinfo {pages} {013010}
  (\bibinfo {year} {2014})}\BibitemShut {NoStop}%
\bibitem [{\citenamefont {Li}\ \emph {et~al.}(2011)\citenamefont {Li},
  \citenamefont {Bensson}, \citenamefont {Nisimova}, \citenamefont {Munger},
  \citenamefont {Mahautmr}, \citenamefont {Tang}, \citenamefont {Maxey},\ and\
  \citenamefont {Brun}}]{Li11}%
  \BibitemOpen
  \bibfield  {author} {\bibinfo {author} {\bibfnamefont {G.}~\bibnamefont
  {Li}}, \bibinfo {author} {\bibfnamefont {J.}~\bibnamefont {Bensson}},
  \bibinfo {author} {\bibfnamefont {L.}~\bibnamefont {Nisimova}}, \bibinfo
  {author} {\bibfnamefont {D.}~\bibnamefont {Munger}}, \bibinfo {author}
  {\bibfnamefont {P.}~\bibnamefont {Mahautmr}}, \bibinfo {author}
  {\bibfnamefont {J.~X.}\ \bibnamefont {Tang}}, \bibinfo {author}
  {\bibfnamefont {M.~R.}\ \bibnamefont {Maxey}}, \ and\ \bibinfo {author}
  {\bibfnamefont {Y.~V.}\ \bibnamefont {Brun}},\ }\href@noop {} {\bibfield
  {journal} {\bibinfo  {journal} {Phys. Rev. E}\ }\textbf {\bibinfo {volume}
  {84}},\ \bibinfo {pages} {041932} (\bibinfo {year} {2011})}\BibitemShut
  {NoStop}%
\bibitem [{\citenamefont {Lee}(2013)}]{Lee2013}%
  \BibitemOpen
  \bibfield  {author} {\bibinfo {author} {\bibfnamefont {C.~F.}\ \bibnamefont
  {Lee}},\ }\href@noop {} {\bibfield  {journal} {\bibinfo  {journal} {New J.
  Phys.}\ }\textbf {\bibinfo {volume} {15}},\ \bibinfo {pages} {055007}
  (\bibinfo {year} {2013})}\BibitemShut {NoStop}%
\bibitem [{\citenamefont {Elgeti}\ and\ \citenamefont
  {Gompper}(2015)}]{Elgeti15}%
  \BibitemOpen
  \bibfield  {author} {\bibinfo {author} {\bibfnamefont {J.}~\bibnamefont
  {Elgeti}}\ and\ \bibinfo {author} {\bibfnamefont {G.}~\bibnamefont
  {Gompper}},\ }\href@noop {} {\bibfield  {journal} {\bibinfo  {journal}
  {Europhys. Lett.}\ }\textbf {\bibinfo {volume} {109}},\ \bibinfo {pages}
  {58003} (\bibinfo {year} {2015})}\BibitemShut {NoStop}%
\bibitem [{\citenamefont {Ezhilan}\ and\ \citenamefont
  {Saintillan}(2015)}]{Ezhilan15}%
  \BibitemOpen
  \bibfield  {author} {\bibinfo {author} {\bibfnamefont {B.}~\bibnamefont
  {Ezhilan}}\ and\ \bibinfo {author} {\bibfnamefont {D.}~\bibnamefont
  {Saintillan}},\ }\href@noop {} {\bibfield  {journal} {\bibinfo  {journal} {J.
  Fluid Mech.}\ }\textbf {\bibinfo {volume} {777}},\ \bibinfo {pages} {482}
  (\bibinfo {year} {2015})}\BibitemShut {NoStop}%
\bibitem [{\citenamefont {Berg}(1993)}]{Berg93}%
  \BibitemOpen
  \bibfield  {author} {\bibinfo {author} {\bibfnamefont {H.~C.}\ \bibnamefont
  {Berg}},\ }\href@noop {} {\emph {\bibinfo {title} {Random Walks in
  Biology}}}\ (\bibinfo  {publisher} {Princeton University Press},\ \bibinfo
  {year} {1993})\BibitemShut {NoStop}%
\bibitem [{\citenamefont {Takatori}\ \emph {et~al.}(2014)\citenamefont
  {Takatori}, \citenamefont {Yan},\ and\ \citenamefont {Brady}}]{Takatori14}%
  \BibitemOpen
  \bibfield  {author} {\bibinfo {author} {\bibfnamefont {S.~C.}\ \bibnamefont
  {Takatori}}, \bibinfo {author} {\bibfnamefont {W.}~\bibnamefont {Yan}}, \
  and\ \bibinfo {author} {\bibfnamefont {J.~F.}\ \bibnamefont {Brady}},\
  }\href@noop {} {\bibfield  {journal} {\bibinfo  {journal} {Phys. Rev. Lett.}\
  }\textbf {\bibinfo {volume} {113}},\ \bibinfo {pages} {028103} (\bibinfo
  {year} {2014})}\BibitemShut {NoStop}%
\bibitem [{\citenamefont {Yang}\ \emph {et~al.}(2014)\citenamefont {Yang},
  \citenamefont {Manning},\ and\ \citenamefont {Marchetti}}]{Yang14}%
  \BibitemOpen
  \bibfield  {author} {\bibinfo {author} {\bibfnamefont {X.}~\bibnamefont
  {Yang}}, \bibinfo {author} {\bibfnamefont {M.~L.}\ \bibnamefont {Manning}}, \
  and\ \bibinfo {author} {\bibfnamefont {M.~C.}\ \bibnamefont {Marchetti}},\
  }\href@noop {} {\bibfield  {journal} {\bibinfo  {journal} {Soft Matter}\
  }\textbf {\bibinfo {volume} {10}},\ \bibinfo {pages} {6477} (\bibinfo {year}
  {2014})}\BibitemShut {NoStop}%
\bibitem [{\citenamefont {Winkler}\ \emph {et~al.}(2015)\citenamefont
  {Winkler}, \citenamefont {Wysocki},\ and\ \citenamefont
  {Gompper}}]{Winkler15}%
  \BibitemOpen
  \bibfield  {author} {\bibinfo {author} {\bibfnamefont {R.~G.}\ \bibnamefont
  {Winkler}}, \bibinfo {author} {\bibfnamefont {A.}~\bibnamefont {Wysocki}}, \
  and\ \bibinfo {author} {\bibfnamefont {G.}~\bibnamefont {Gompper}},\
  }\href@noop {} {\bibfield  {journal} {\bibinfo  {journal} {submitted}\ }
  (\bibinfo {year} {2015})}\BibitemShut {NoStop}%
\bibitem [{\citenamefont {Solon}\ \emph
  {et~al.}(2015{\natexlab{a}})\citenamefont {Solon}, \citenamefont
  {Stenhammar}, \citenamefont {Wittkowski}, \citenamefont {Kardar},
  \citenamefont {Kafri}, \citenamefont {Cates},\ and\ \citenamefont
  {Tailleur}}]{Solon15}%
  \BibitemOpen
  \bibfield  {author} {\bibinfo {author} {\bibfnamefont {A.~P.}\ \bibnamefont
  {Solon}}, \bibinfo {author} {\bibfnamefont {J.}~\bibnamefont {Stenhammar}},
  \bibinfo {author} {\bibfnamefont {R.}~\bibnamefont {Wittkowski}}, \bibinfo
  {author} {\bibfnamefont {M.}~\bibnamefont {Kardar}}, \bibinfo {author}
  {\bibfnamefont {Y.}~\bibnamefont {Kafri}}, \bibinfo {author} {\bibfnamefont
  {M.~E.}\ \bibnamefont {Cates}}, \ and\ \bibinfo {author} {\bibfnamefont
  {J.}~\bibnamefont {Tailleur}},\ }\href@noop {} {\bibfield  {journal}
  {\bibinfo  {journal} {Phys. Rev. Lett.}\ }\textbf {\bibinfo {volume} {114}},\
  \bibinfo {pages} {198301} (\bibinfo {year} {2015}{\natexlab{a}})}\BibitemShut
  {NoStop}%
\bibitem [{\citenamefont {Takatori}\ and\ \citenamefont
  {Brady}(2015)}]{Takatori15}%
  \BibitemOpen
  \bibfield  {author} {\bibinfo {author} {\bibfnamefont {S.~C.}\ \bibnamefont
  {Takatori}}\ and\ \bibinfo {author} {\bibfnamefont {J.~F.}\ \bibnamefont
  {Brady}},\ }\href@noop {} {\bibfield  {journal} {\bibinfo  {journal} {Phys.
  Rev. E}\ }\textbf {\bibinfo {volume} {91}},\ \bibinfo {pages} {032117}
  (\bibinfo {year} {2015})}\BibitemShut {NoStop}%
\bibitem [{\citenamefont {Ray}\ \emph {et~al.}(2014)\citenamefont {Ray},
  \citenamefont {Reichhardt},\ and\ \citenamefont {{Olson
  Reichhardt}}}]{Ray14}%
  \BibitemOpen
  \bibfield  {author} {\bibinfo {author} {\bibfnamefont {D.}~\bibnamefont
  {Ray}}, \bibinfo {author} {\bibfnamefont {C.}~\bibnamefont {Reichhardt}}, \
  and\ \bibinfo {author} {\bibfnamefont {C.~J.}\ \bibnamefont {{Olson
  Reichhardt}}},\ }\href@noop {} {\bibfield  {journal} {\bibinfo  {journal}
  {Phys. Rev. E}\ }\textbf {\bibinfo {volume} {90}},\ \bibinfo {pages} {013019}
  (\bibinfo {year} {2014})}\BibitemShut {NoStop}%
\bibitem [{\citenamefont {Mallory}\ \emph {et~al.}(2014)\citenamefont
  {Mallory}, \citenamefont {{Sari\'c}}, \citenamefont {Valeriani},\ and\
  \citenamefont {Cacciuto}}]{Mallory14}%
  \BibitemOpen
  \bibfield  {author} {\bibinfo {author} {\bibfnamefont {S.~A.}\ \bibnamefont
  {Mallory}}, \bibinfo {author} {\bibfnamefont {A.}~\bibnamefont {{Sari\'c}}},
  \bibinfo {author} {\bibfnamefont {C.}~\bibnamefont {Valeriani}}, \ and\
  \bibinfo {author} {\bibfnamefont {A.}~\bibnamefont {Cacciuto}},\ }\href@noop
  {} {\bibfield  {journal} {\bibinfo  {journal} {Phys. Rev. E}\ }\textbf
  {\bibinfo {volume} {89}},\ \bibinfo {pages} {052303} (\bibinfo {year}
  {2014})}\BibitemShut {NoStop}%
\bibitem [{\citenamefont {Ginot}\ \emph {et~al.}(2015)\citenamefont {Ginot},
  \citenamefont {Theurfauff}, \citenamefont {Levis}, \citenamefont {Ybert},
  \citenamefont {Bocquet}, \citenamefont {Berthier},\ and\ \citenamefont
  {{Cottin-Bizonne}}}]{Ginot15}%
  \BibitemOpen
  \bibfield  {author} {\bibinfo {author} {\bibfnamefont {F.}~\bibnamefont
  {Ginot}}, \bibinfo {author} {\bibfnamefont {I.}~\bibnamefont {Theurfauff}},
  \bibinfo {author} {\bibfnamefont {D.}~\bibnamefont {Levis}}, \bibinfo
  {author} {\bibfnamefont {C.}~\bibnamefont {Ybert}}, \bibinfo {author}
  {\bibfnamefont {L.}~\bibnamefont {Bocquet}}, \bibinfo {author} {\bibfnamefont
  {L.}~\bibnamefont {Berthier}}, \ and\ \bibinfo {author} {\bibfnamefont
  {C.}~\bibnamefont {{Cottin-Bizonne}}},\ }\href@noop {} {\bibfield  {journal}
  {\bibinfo  {journal} {Phys. Rev. X}\ }\textbf {\bibinfo {volume} {5}},\
  \bibinfo {pages} {011004} (\bibinfo {year} {2015})}\BibitemShut {NoStop}%
\bibitem [{\citenamefont {Solon}\ \emph
  {et~al.}(2015{\natexlab{b}})\citenamefont {Solon}, \citenamefont {Fily},
  \citenamefont {Baskaran}, \citenamefont {Cates}, \citenamefont {Kafri},
  \citenamefont {Kardar},\ and\ \citenamefont {Tailleur}}]{Solon15b}%
  \BibitemOpen
  \bibfield  {author} {\bibinfo {author} {\bibfnamefont {A.~P.}\ \bibnamefont
  {Solon}}, \bibinfo {author} {\bibfnamefont {Y.}~\bibnamefont {Fily}},
  \bibinfo {author} {\bibfnamefont {A.}~\bibnamefont {Baskaran}}, \bibinfo
  {author} {\bibfnamefont {M.~E.}\ \bibnamefont {Cates}}, \bibinfo {author}
  {\bibfnamefont {Y.}~\bibnamefont {Kafri}}, \bibinfo {author} {\bibfnamefont
  {M.}~\bibnamefont {Kardar}}, \ and\ \bibinfo {author} {\bibfnamefont
  {J.}~\bibnamefont {Tailleur}},\ }\href@noop {} {\bibfield  {journal}
  {\bibinfo  {journal} {Nature Phys.}\ ,\ \bibinfo {pages}
  {DOI:10.1038/NPHYS3377}} (\bibinfo {year} {2015}{\natexlab{b}})}\BibitemShut
  {NoStop}%
\bibitem [{\citenamefont {Fily}\ \emph {et~al.}(2014)\citenamefont {Fily},
  \citenamefont {Baskaran},\ and\ \citenamefont {Hagan}}]{Fily14}%
  \BibitemOpen
  \bibfield  {author} {\bibinfo {author} {\bibfnamefont {Y.}~\bibnamefont
  {Fily}}, \bibinfo {author} {\bibfnamefont {A.}~\bibnamefont {Baskaran}}, \
  and\ \bibinfo {author} {\bibfnamefont {M.~F.}\ \bibnamefont {Hagan}},\
  }\href@noop {} {\bibfield  {journal} {\bibinfo  {journal} {Soft Matter}\
  }\textbf {\bibinfo {volume} {10}},\ \bibinfo {pages} {5609} (\bibinfo {year}
  {2014})}\BibitemShut {NoStop}%
\bibitem [{\citenamefont {Kantsler}\ \emph {et~al.}(2013)\citenamefont
  {Kantsler}, \citenamefont {Dunkel}, \citenamefont {Polin},\ and\
  \citenamefont {Goldstein}}]{Kantsler13}%
  \BibitemOpen
  \bibfield  {author} {\bibinfo {author} {\bibfnamefont {V.}~\bibnamefont
  {Kantsler}}, \bibinfo {author} {\bibfnamefont {J.}~\bibnamefont {Dunkel}},
  \bibinfo {author} {\bibfnamefont {M.}~\bibnamefont {Polin}}, \ and\ \bibinfo
  {author} {\bibfnamefont {R.~E.}\ \bibnamefont {Goldstein}},\ }\href@noop {}
  {\bibfield  {journal} {\bibinfo  {journal} {Proc. Natl. Acad. Sci. USA}\
  }\textbf {\bibinfo {volume} {110}},\ \bibinfo {pages} {1187} (\bibinfo {year}
  {2013})}\BibitemShut {NoStop}%
\bibitem [{\citenamefont {Volpe}\ \emph {et~al.}(2011)\citenamefont {Volpe},
  \citenamefont {Buttinoni}, \citenamefont {Vogt}, \citenamefont
  {{K\"ummerer}},\ and\ \citenamefont {Bechinger}}]{Volpe11}%
  \BibitemOpen
  \bibfield  {author} {\bibinfo {author} {\bibfnamefont {G.}~\bibnamefont
  {Volpe}}, \bibinfo {author} {\bibfnamefont {I.}~\bibnamefont {Buttinoni}},
  \bibinfo {author} {\bibfnamefont {D.}~\bibnamefont {Vogt}}, \bibinfo {author}
  {\bibfnamefont {H.-J.}\ \bibnamefont {{K\"ummerer}}}, \ and\ \bibinfo
  {author} {\bibfnamefont {C.}~\bibnamefont {Bechinger}},\ }\href@noop {}
  {\bibfield  {journal} {\bibinfo  {journal} {Soft Matter}\ }\textbf {\bibinfo
  {volume} {7}},\ \bibinfo {pages} {8810} (\bibinfo {year} {2011})}\BibitemShut
  {NoStop}%
\end{thebibliography}%

\end{document}